\documentclass[a4paper,11pt]{article}
\usepackage{pos}
\usepackage{slashed}

\title{Two-loop master integrals for a planar topology contributing to $pp \rightarrow t\bar{t}j$}

\author*[a]{Matteo Becchetti}

\affiliation[a]{Dipartimento di Fisica e Astronomia, Università di Bologna e INFN, Sezione di Bologna, via Irnerio 46, I-40126 Bologna, Italy}

\emailAdd{matteo.becchetti@unibo.it}

\abstract{We report on recent progress for the QCD corrections to top quark pair plus jet production. In particular, we discuss a recent computation for the two-loop master integrals associated to a two-loop five-point pentagon-box integral configuration with one internal massive propagator, that contributes to top quark pair production in association with a jet in the QCD planar limit.}

\FullConference{16th International Symposium on Radiative Corrections: Applications of Quantum Field Theory to Phenomenology (
RADCOR2023)\\
28th May - 2nd June, 2023\\
Crieff, Scotland, UK\\}

\newcommand{\beq}{\begin{equation}}

\newcommand{\eeq}{\end{equation}}
\newcommand{\nn}{\nonumber}
\newcommand{\bea}{\begin{eqnarray}}
\newcommand{\eea}{\end{eqnarray}}
\newcommand{\bfig}{\begin{figure}}
\newcommand{\efig}{\end{figure}}
\newcommand{\bc}{\begin{center}}
\newcommand{\ec}{\end{center}}

\newcommand{\eps}{{\varepsilon}}
\newcommand{\tb}{{\bar{t}}}

\definecolor{mypink}{RGB}{219, 48, 122}
\definecolor{mygreen}{rgb}{0,0.7,0}
\definecolor{raspberry}{rgb}{0.53,0.15,0.34}

\begin{document}
\maketitle

\section{Introduction}

The Large Hadron Collider (LHC) is entering the high-precision era with the High-Luminosity plan 
(HI-LHC). This project will enable experimental collaborations to measure many interesting observables
at percent level precision. In order to be able to compare the experimental measurements with
theoretical predictions, it is mandatory to achieve a theoretical uncertainty on the same level
of the experimental one. One of the ingredients that are needed in order to achieve this goal
are next-to-next-to leading order (NNLO) QCD corrections. While a lot progress has been done
recently in this framework, QCD corrections at NNLO are still not available for all the most
interesting observables at LHC.

One of the observables for which NNLO QCD corrections are yet to be obtained is the top quark
pair production in association with a jet. As the top quark is the heaviest particle in the Standard Model (SM) of particle physics, 
it has many important implications for the nature of the fundamental
forces. In particular many properties of the SM are sensitive to the value of the top quark mass as,
for example, the stability of the SM vacuum whose precision measurement is a high priority at the (LHC).
The standard process which is exploited to measure the top quark mass at the LHC is top quark pair production. 
This process is known with vey high precision both theoretically and experimentally~\cite{Czakon:2019yrx,Cooper-Sarkar:2020twv}.
However, it has been recently argued that top quark pair production in association with a jet is even more sensitive to
 the top quark mass~\cite{Alioli:2013mxa,Bevilacqua:2017ipv,Alioli:2022lqo}. The state-of-the-art for the
theoretical predictions of this process is represented by the next-to-leading order (NLO) QCD
corrections~\cite{Dittmaier:2007wz,Dittmaier:2008uj}, along with complete decay information and interfaces with a parton
shower~\cite{Melnikov:2010iu,Alioli:2011as,Czakon:2015cla,Bevilacqua:2015qha,Bevilacqua:2016jfk}.
However, in order to match the experimental precision, see for
example~\cite{ATLAS:2019guf,CMS:2020grm}, next-to-next-to-leading order (NNLO)
corrections are required.

In order to be able to perform a full NNLO prediction for this observable
several computational difficulties have to be overcome.
One of the major obstacles is the computation of the required two-loop scattering amplitudes.
Recently a great progress has been made in the calculation of scattering amplitudes for
$2\to3$ processes \cite{Gehrmann:2015bfy,Badger:2018enw,Abreu:2018aqd,Chicherin:2018yne,Abreu:2018zmy,Abreu:2019odu,Abreu:2018jgq,Badger:2019djh,Abreu:2020cwb,Abreu:2021oya,Chawdhry:2020for,Hartanto:2019uvl,Agarwal:2021grm,Chawdhry:2021mkw,Agarwal:2021vdh,Badger:2021imn,Badger:2021ega,Badger:2022ncb,Badger:2023xtl,Zoia:2023nup}, which led to
a number of NNLO QCD theoretical predictions \cite{Chawdhry:2019bji,Kallweit:2020gcp,Chawdhry:2021hkp,Badger:2021ohm,Hartanto:2022qhh,Badger:2023mgf}.
Yet, the amplitudes necessary to perform 
a NNLO theoretical prediction for top quark pair plus a jet production at LHC represent
a substantial step forward with respect to the current state-of-the-art.
Indeed, the top quark mass which appear in the internal propagators is responsible
for a significant growth in the complexity of the computation. This feature affect
both the algebraic complexity in the amplitude reconstruction, and the analytic
complexity of the Feynman integrals.

In this context, I report on the recent progress made in the computation of two-loop Feynman integrals
relevant for the NNLO QCD corrections to $pp\to t\tb j$ \cite{Badger:2022hno}. This project builds upon previous work where the
authors computed the one-loop helicity amplitudes expanded up to $\mathcal{O}(\eps^2)$ in the
dimensional regulator \cite{Badger:2022mrb}, which are needed for the NNLO correections.
 In \cite{Badger:2022hno} the authors
studied the master integrals associated to a five-point pentagon-box topology with one internal
massive propagator, that contributes to top-quark pair production in association with a jet in the leading color QCD planar limit.
The computation represents a step forward in complexity with respect to the five-point massless \cite{Gehrmann:2015bfy,Papadopoulos:2015jft,Gehrmann:2018yef,Abreu:2018aqd,Chicherin:2018old,Chicherin:2020oor} and one off-shell
external leg cases \cite{Abreu:2020jxa,Canko:2020ylt,Chicherin:2021dyp,Abreu:2021smk,Abreu:2023rco}. 

The master integrals have been computed exploiting the differential equation method \cite{Kotikov:1990kg,Remiddi:1997ny}.
The system of differential equations has been written with respect to a canonical basis of master 
integrals \cite{Henn:2013pwa}. A major bottleneck for this computation is the solution
of a large system of Integration-by-Parts (IBP)
relations \cite{Tkachov:1981wb,Chetyrkin:1981qh}. In order to overcome this issue finite
fields arithmetic \cite{vonManteuffel:2014ixa,Peraro:2016wsq,Peraro:2019svx}, as
implemented in the \textsc{FiniteFlow} library~\cite{Peraro:2019svx}, has been employed. We obtained
a semi-analytic solution for the master integrals through the generalised power series method
\cite{Lee:2017qql,Mandal:2018cdj,Francesco:2019yqt}, as described in \cite{Francesco:2019yqt} 
and implemented in the \textsc{Mathematica} package \textsc{DiffExp}~\cite{Hidding:2020ytt}.
In order to solve the system of differential equations semi-analytically, we used high precision
numerical boundary conditions obtained by means of the \textsc{Mathematica}
package \textsc{AMFlow}~\cite{Liu:2022chg}, which implements the auxiliary mass flow method
\cite{Liu:2017jxz,Liu:2021wks,Liu:2022tji}. Finally, we also derived the
analytic representation of the alphabet for the system of differential equations. Interestingly, 
the structure of the alphabet has the same
analytic structure as in the five-point massless \cite{Gehrmann:2015bfy} and in
the one-mass \cite{Abreu:2020jxa} cases.

The outcome of the work presented in \cite{Badger:2022hno}, and summarised in the present
proceeding, is two-fold. First, we obtained a solution for the master integrals under study
which has the potential for phenomenological applications, as it has been done for other processes
\cite{Bonciani:2016qxi,Bonciani:2019jyb,Frellesvig:2019byn,Abreu:2020jxa,Becchetti:2020wof,Abreu:2021smk,Armadillo:2022bgm,Bonciani:2021zzf,Badger:2022mrb,Becchetti:2023yat}.
Moreover, the study of the analytic structure of the alphabet solution is a fundamental step in order to achieve
a complete analytic representation. As a consequence, the work presented in \cite{Badger:2022hno} represents
a fist step toward an analytic computation for the NNLO QCD corrections to top quark pair production
in association with a jet in the QCD leading color planar limit.

\section{Summary of the computation}

\begin{figure}[t!]
\begin{center}
\includegraphics[width = 0.4\linewidth]{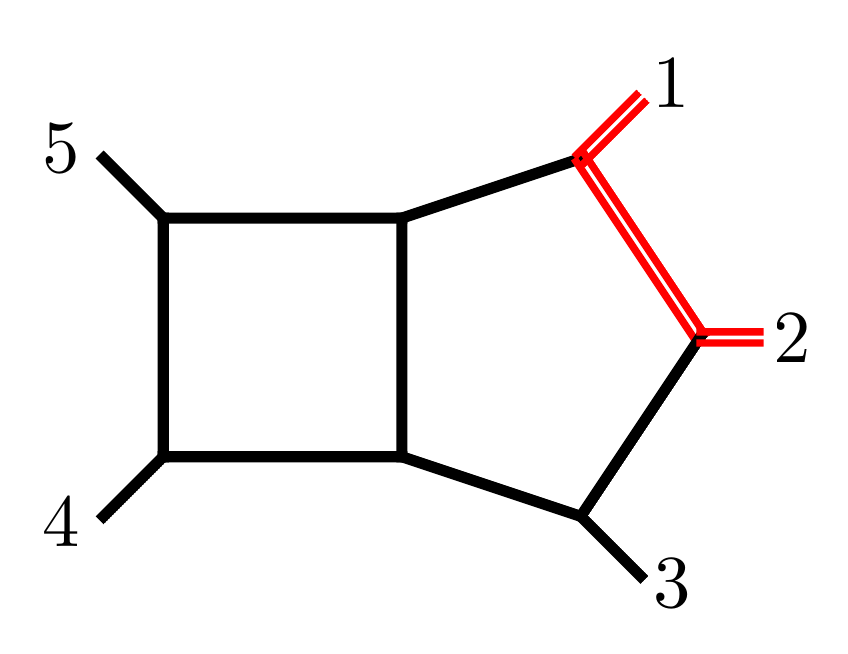}
\end{center}
\caption{The pentagon-box topology contributing to  $pp \rightarrow t\bar{t}j$ in the QCD leading color planar limit. Black lines denote massless particles and red double-lines denote massive particles.}
\label{fig:t431def}
\end{figure}

We considered the pentagon-box Feynman integral topology in $d=4-2\eps$ dimensions as shown in figure \ref{fig:t431def}. This can be written as,
\begin{equation}
  I_{a_1,a_2,a_3,a_4,a_5,a_6,a_7,a_8}^{a_9,a_{10},a_{11}} = \int \mathcal{D}^{d} k_1 \mathcal{D}^{d} k_2 
  \frac{D_{9}^{a_9} D_{10}^{a_{10}} D_{11}^{a_{11}}}{D_{1}^{a_1}\cdots D_{8}^{a_8}}\,
  \label{eq:t431def}
\end{equation}
where $a_1,\cdots,a_{11} \geq 0$. The topology is defined by the following set of propagators, and numerators:
\begin{align}
  D_1 & =  k_1^2, & D_2 &= (k_1 - p_1)^2 - m_t^2, & D_3 &= (k_1 - p_1 - p_2)^2,  \nonumber \\ 
  D_4 &= (k_1 - p_1 - p_2 - p_3)^2, & D_5 &= k_2^2, & D_6  & =  (k_2 - p_5)^2, \nonumber \\
  D_7 &= (k_2 - p_4 - p_5)^2, & D_8 &= (k_1 +k_2)^2,    &
  D_9 &= (k_1 + p_5)^2, \nonumber \\ \label{eq:props}
  D_{10} &  =   (k_2 + p_1)^2 -m_t^2, &  D_{11} &= (k_2 + p_1 + p_2)^2,& & 
\end{align}
and the integration measure is:
\begin{equation}
  \mathcal{D}^d k_i = \dfrac{d^d k_i}{i \pi^{\frac{d}{2}}} e^{\eps \gamma_E}  \,.
\end{equation}
The momenta are considered outgoing from the graphs and the particles are on-shell, i.e. $p_1^2 = p_2^2 = m_t^2$ for the top quark external legs, while $p_3^2 = p_4^2 =
p_5^2=0$. The kinematics of the integrals is described by six
independent invariants $\vec{x}=\{d_{12}, d_{23}, d_{34}, d_{45}, d_{15}, m_t^2\}$, where
\begin{equation}
  d_{ij} = p_i \cdot p_j,
  \label{eq:dijdef}
\end{equation}
and $m_t^2$ is the top quark squared mass.
After performing IBP reduction \cite{Chetyrkin:1981qh,Chetyrkin:1979bj}, as implemented in the software \textsc{LiteRed} \cite{Lee:2012cn,Lee:2013mka}
and \textsc{FiniteFlow} \citep{Peraro:2019svx}, we found a total number of 88 MIs (see \cite{Badger:2022hno} for the complete list).

We wrote a system of differential equations for the MIs $\vec{\mathcal{I}}(\vec{x},\eps)$ in canonical form \cite{Henn:2013pwa}:
\begin{equation} \label{eq:deqsCan}
 d\, \vec{\mathcal{I}}(\vec{x},\eps) = \eps \, d A(\vec{x}) \, \vec{\mathcal{I}}(\vec{x},\eps),
\end{equation}   
where $d$ is the total differential with respect to the kinematic invariants, and the matrix $A(\vec{x})$ is a linear combination of logarithms:
\begin{equation}\label{eq:dematrix}
 A(\vec{x}) = \sum c_i \log (w_i (\vec{x})).
\end{equation}
The $c_i$ are matrices of rational numbers, and the \emph{alphabet} $\left\{w_i (\vec{x})\right\}$ is made by
algebraic functions of the kinematic invariants $\vec{x}$.

\subsection{Canonical Basis}

The canonical basis of MIs $\vec{\mathcal{I}}$ has been constructed starting from the observation
of an emerging pattern for $2\to 3$ scattering amplitudes \cite{Gehrmann:2015bfy,Papadopoulos:2015jft,Abreu:2018rcw,Chicherin:2018yne,Chicherin:2019xeg,Chicherin:2018old,Abreu:2018aqd,Abreu:2019rpt,Abreu:2020jxa,Abreu:2021smk,Canko:2020ylt}. This feature implies that we are able to rely on a good set of  uniform trascendental (UT) candidate MIs
as starting point for the basis construction. Specifically, one can test candidates 
from the MIs basis for the massless and one-mass five-point \cite{Gehrmann:2018yef,Chicherin:2018old,Abreu:2020jxa,Abreu:2021smk} (for e.g.
$pp\to W+2j$ and $pp\to 3j$) cases, as well as other integral topologies
which involve internal massive propagators\footnote{For example a large number of MIs for the two-mass four-point $pp\to Wt$ scattering~\cite{Chen:2021gjv} appear as subtopologies in our 88 integral system. This feature allowed us to reduce the number of completely unknown MIs in UT form to 40.}.

Guided by this initial set of data, our approach relies on the possibility to perform IBP reduction and evaluate the
differential equations matrix over finite fields. Due to the presence of square roots, we do not attempt to construct
the canonical form directly. Instead we search for a linear form, with respect to $\eps$, which contains only 
rational matrices in the kinematic invariants. Indeed, the square roots appearing in the UT basis can be
absorbed in the normalisation of the integral basis\footnote{This approach is discussed in Ref.
\cite{Peraro:2019svx}.}, and therefore we can neglect them while evaluating the differential equations
over finite fields.
The strategy adopted in \cite{Badger:2022hno} can be then summarised as follows:
\begin{itemize}
\item Given a starting set of UT candidate MIs, we study the $\eps$ structure of the differential
equations from a univariate slice reconstruction. Specifically, we search for a linear form in $\eps$
\begin{equation}
  d\, \vec{\mathcal{J}}(\vec{x},\eps) = \, d \, \left(\hat{A}^{(0)}(\vec{x}) + \eps \widehat{A}^{(1)}(\vec{x}) \right) \, \vec{\mathcal{J}}(\vec{x},\eps),
  \label{eq:deqsCan2}
\end{equation}
where $\widehat{A}^{(0)}$ is a diagonal matrix;   
\item We study the homogenous part of the system of differential equations sector-by-sector, in order to determine
the correct normalisation for the MIs;
\item If the starting choice of integral basis, for a given sector, does not satisfies a differential equations of the form in
Eq. \eqref{eq:deqsCan2}, we make a different ansatz based on criteria described below. 
\end{itemize}
Once the whole system of differential equations is in the form of Eq. \eqref{eq:deqsCan2} we can rotate it into canonical form:
\begin{equation}
  \mathcal{I}_i = N_{ij}(\vec{x}) \mathcal{J}_j ,
  \label{eq:MISnorm}
\end{equation}
where $N_{ij}(\vec{x})$ is a diagonal matrix which contains the square roots of the kinematic invariants. Such matrix satisfies
the differential equation:
\begin{equation}
  \widehat{A}^{(0)} - \frac{1}{2} N^2  d N^{(-2)}  = 0.
\end{equation}
The canonical form of the differential equations can then be written as:
\begin{equation}
  d\, \vec{\mathcal{I}}(\vec{x},\eps) = \eps d\left( N(\vec{x})\widehat{A}^{(1)}(\vec{x})N^{-1}(\vec{x}) \right) \, \vec{\mathcal{I}}(\vec{x},\eps)
\end{equation}

As anticipated, if the starting integral basis does not satisfies a differential equations of the form in Eq. \eqref{eq:deqsCan2} we 
change the starting ansatz. This is done accordingly to a set of criteria inspired by patterns observed in previously studied cases: 
\begin{itemize}
  \item For two and three external legs MIs the choice of candidates can involve scalar integrals with dotted denominators;
  \item For four external legs MIs the choice of candidates can involve scalar integrals with dotted denominators or the numerators $D_9,D_{10},D_{11}$;
  \item For five external legs, canonical MIs candidates can involve scalar integrals with the numerators $D_9,D_{10},D_{11}$ and local integrand insertions $\mu_{ij}$,
\end{itemize}
where $\mu_{ij}$ are defined from the splitting of the loop momenta into four dimensional and $(-2\eps)$ dimensional components,
\begin{align}
  k_i &= k_i^{[4]} + k_i^{[-2\eps]}, & \mu_{ij} &= -k_i^{[-2\epsilon]}\cdot k_j^{[-2\epsilon]}.
\end{align}

I conclude the present discussion with some remarks. First, given the high number of kinematic
invariants, and the large size of the IBPs systems to solve, it is important to ensure that the
maximum numerator rank and number of dotted propagators is minimised. 
As second remark, I mention that the method exploited to build a canonical basis might still require some
work on the sub-topologies contribution to the differential equations for
a given sector. Indeed, we found that some sectors required additional rotations in
sub-sectors. However, this step was particularly simple in our cases. 
Interestingly, such feature did not
appear in any of the most complicated five-point topologies, where the
UT integrals can be constructed exploiting just local numerator insertions.

\subsection{Analytic structure}

Even though the system of differential equations has been integrate semi-analytically
exploiting the generalised series expansion method, we studied the alphabet structure of the
solution. This aspect is crucial for understanding the analytic solution and it is 
the first step towards constructing a well defined special function basis for the set of MIs
under consideration.

The system of differential equations can be written in terms of d-logarithmic forms using an alphabet which is made of 71 letters $w_i$:
\begin{equation}
 d\, \vec{\mathcal{I}}(\vec{x},\eps) = \eps \, d A(\vec{x}) \, \vec{\mathcal{I}}(\vec{x},\eps), \,\,\, A(\vec{x}) = \sum_{i = 1}^{71} c_i \log (w_i (\vec{x})).
\end{equation}
In order to identify the alphabet we adopted a
strategy along the lines of the one described in
Refs. \cite{Heller:2019gkq,Zoia:2021zmb,Chaubey:2022hlr}, which we briefly
summarise. As first step we identify the set of rational letters
inside the alphabet. This can be done by looking at the denominators
in the differential equations system. The remaining letters are, therefore, algebraic in the
kinematic invariants (i.e. they contain square roots). To obtain
this set of letters we proceed as follows. We determine the linear relations 
in the total derivative matrix $d A(\vec{x})$ and we find a minimal
set of independent one forms. Then, for each independent entry of
the derivative matrix one determines which square roots appear in the
denominators. Finally, it is possible to construct an ansatz using free polynomials in
the variables $d_{ij}$ which depends on the square roots in the one-form under study. 
The form of the ansatz depends on the number of square roots, e.g. if there is just one square root
we can use an ansatz of the kind,
\begin{equation}
  \Omega(a,b) := \frac{a + \sqrt{b}}{a - \sqrt{b}},
  \label{eq:lettertemplate1}
\end{equation}
and in the case of two square roots,
\begin{equation}
  \tilde{\Omega}(a,b,c) := \frac{(a + \sqrt{b} + \sqrt{c})(a - \sqrt{b} - \sqrt{c})}{(a + \sqrt{b} - \sqrt{c} )(a -\sqrt{b} + \sqrt{c} )}.
  \label{eq:lettertemplate2}
\end{equation}
While it is always possible to expand the form of Eq. \eqref{eq:lettertemplate2} into one
similar to Eq. \eqref{eq:lettertemplate1}, the structure in Eq. \eqref{eq:lettertemplate2} is
preferable. Indeed, the polynomial degree of the unknown variable $a$ is lower as
noted in Ref.~\cite{Abreu:2020jxa}.

Following this strategy we have identified an alphabet which can be split into two subsets, $\mathbf{W}_R$ and $\mathbf{W}_A$, which are, respectively, rational and
algebraic in the kinematic invariants. For the rational letters we define,
\begin{equation}
\mathbf{W}_{R} := \mathbf{W}_K \cup \mathbf{W}_T \cup \mathbf{W}_S,
\end{equation}
and for the algebraic letters
\begin{equation}
\mathbf{W}_{A} := \mathbf{W}_{SR-1} \cup \mathbf{W}_{TR} \cup \mathbf{W}_{SR-2}.
\end{equation}
The rational set of letters $\mathbf{W}_R$ can be furthermore divided into three subsets. The subset $\mathbf{W}_K$ is made by letters which are linear combinations of the Mandelstam variables $s_{ij} = (p_i + p_j)^2$. The letters in the subset $\mathbf{W}_T$ can be written as traces over $\gamma$-matrices:
\begin{align}
  \operatorname{tr}(ij\cdots k) &= \operatorname{tr}(\slashed{p}_i\slashed{p}_j \cdots \slashed{p}_k).
\end{align}
Finally, the rational letters in the third subset, $\mathbf{W}_S$, are related to the roots that appear in the differential equations system:
\begin{align}
\mathbf{W}_S := & \left\{\beta^2, \, (\Delta_1)^2, \, (\Delta_2)^2, \,4 (d_{12}+d_{23}+m_t^2)^2 (\Delta_3)^2, \, (\Delta_5)^2, \, (\Delta_4)^2, \, (\Delta_6)^2, \, \operatorname{tr}_5^2\right\},
\end{align}
which are defined as follows:
\begin{align}
  \beta & = \sqrt{1-\frac{4 m_t^2}{s_{12}}},  \nn \\
  \Delta_1 &= \sqrt{\operatorname{det}G(p_{23},p_1)},  && \Delta_2 = \sqrt{\operatorname{det}G(p_{15},p_2)}, \nn \\
  \Delta_3 & = \sqrt{1 - \frac{4 s_{45} m_t^2}{(s_{12}+s_{23}-m_t^2)^2}},  && \Delta_4 = \sqrt{1 + \frac{4 s_{34} s_{45} m_t^2}{s_{12} (s_{15}-s_{23})^2}}, \nn \\
  \Delta_5 & = \sqrt{1-\frac{s_{45} m_t^2}{4 d_{15} d_{23}}},  && \Delta_6 = \sqrt{1-\frac{s_{34} s_{45} m_t^2}{4 d_{15} d_{23} s_{12}}}, \nn \\
  \operatorname{tr}_5 & = 4 \sqrt{\operatorname{det}G(p_3,p_4,p_5,p_1)} = {\rm tr}(\gamma_5 \slashed{p}_3 \slashed{p}_4 \slashed{p}_5 \slashed{p}_1), 
  \label{eq:sqrt}
\end{align}
where $G_{ij}(\vec{v}) = v_i\cdot v_j $ is the Gram matrix.

Similarly to the rational subset of letters, also the algebraic one $\mathbf{W}_A$ can be split into three subsets.  The first one,
$\mathbf{W}_{SR-1}$, contains letters which can be written in terms of the quantity $\Omega$ as defined above in Eq. \eqref{eq:lettertemplate1}. 
Instead, the letters associated to the subset $\mathbf{W}_{TR}$, contain dependence on the Dirac $\gamma_5$ matrix.
Therefore, they can be written as ratios of $ \operatorname{tr}_{\pm}(ij\cdots k)$ objects, defined as
\begin{equation}
  \operatorname{tr}_{\pm}(ij\cdots k) = \frac{1}{2}\operatorname{tr}((1 \pm \gamma_5)\slashed{p}_i\slashed{p}_j \cdots \slashed{p}_k).
  \label{eq:lettertemplate3}
\end{equation}
The final subset, $\mathbf{W}_{SR-2}$, is made by letters in terms of $\tilde{\Omega}$ as defined above in Eq. \eqref{eq:lettertemplate2}. 

I finish this discussion we the following consideration. The alphabet structure just presented shows a similar pattern to the ones observed in other five-particle kinematic configurations \cite{Chicherin:2017dob,Gehrmann:2018yef,Abreu:2020jxa,Abreu:2021smk}.
This feature suggests that it might exists a general alphabet structure for all polylogarihmic two-loop integrals with five or fewer legs.

\subsection{Numerical Evaluation}

In order to validate our work we exploited the package \textsc{DiffExp} \cite{Hidding:2020ytt}
to evaluate numerically the MIs. This package implements the generalised power
series method \cite{Francesco:2019yqt}, which gives a semi-analytical solution
to the system of differential equations as an expansion around its singular
points:
\begin{equation}
\vec{\mathcal{I}}(t, \epsilon) = \sum_{k=0}^{\infty} \epsilon^k \sum_{i = 0}^{N - 1}\rho_i (t) \vec{\mathcal{I}}^{(k)}_i (t), \;\;\; \rho(t)  = 
\begin{cases}
1, & t \in \left[t_i -r_i, t_i + r_i\right) \\
0, & t \notin \left[t_i -r_i, t_i + r_i\right)
\end{cases}, 
\end{equation}
\begin{equation}
\vec{\mathcal{I}}_i^{(k)}(t) = \sum_{l_1 = 0}^{\infty}\sum_{l_2 = 0}^{N_{i,k}} c_k^{(i,l_1,l_2)} \left(t - t_i\right)^{\frac{l_1}{2}} \log(t-t_i)^{l_2}.
\end{equation}
In the previous equations $t$ is a variable that parametrise the integration path
in the kinematic invariants space, $t_i$ are singular points for the system of
differential equations, $r_i$ is the radius of convergence of the series solution
around $t_i$ and $c_k^{(i,l_1,l_2)}$ are matrices which depend on the system
of differential equations and the boundary conditions. Since we were interested in a
numerical evaluation of the MIs, the system of differential equations has been integrated
using high-precision numerical boundary conditions obtained with the package
\textsc{AMFlow} \cite{Liu:2022chg}, which implements the auxiliary mass flow method  \cite{Liu:2017jxz,Liu:2021wks,Liu:2022tji}.
The numerical solution obtained with \textsc{DiffExp} has been checked for
several points against an independent evaluation performed with \textsc{AMFlow}
finding full agreement for all the points under study.

The solution for the MIs presented in Ref. \cite{Badger:2022hno} has not been optimised
for a realistic phase-space integration. However, the successful applications of the
generalised power series method to phenomenological studies in Refs. \cite{Bonciani:2016qxi,Bonciani:2019jyb,Frellesvig:2019byn,Abreu:2020jxa,Becchetti:2020wof,Abreu:2021smk,Armadillo:2022bgm,Bonciani:2021zzf,Badger:2022mrb,Becchetti:2023yat},
offers hope that a phenomenologically oriented improvement of the implementation previously discussed may be achievable
in the near future. 

\section{Acknowledgments}

I thank Simon Badger, Ekta Chaubey and Robin Marzucca as co-authors of the work in Ref. \cite{Badger:2022hno} on which
this proceeding is based on. This project received funding from the European Union's Horizon 2020 research and innovation programmes \textit{High precision
multi-jet dynamics at the LHC} (consolidator grant agreement No 772099).

\bibliographystyle{JHEP}
\bibliography{ppttj_431}

\end{document}